\documentclass[lettersize,journal]{IEEEtran}
\usepackage{amsmath,amsfonts}
\usepackage{algorithmic}
\usepackage{array}
\usepackage[caption=false,font=normalsize,labelfont=sf,textfont=sf]{subfig}
\usepackage{textcomp}
\usepackage{stfloats}
\usepackage{amssymb}
\usepackage{url}
\usepackage{verbatim}
\usepackage{graphicx}
 \usepackage{booktabs} 
\def\BibTeX{{\rm B\kern-.05em{\sc i\kern-.025em b}\kern-.08em
    T\kern-.1667em\lower.7ex\hbox{E}\kern-.125emX}}
\usepackage{balance}

\begin{document}

\title{HTS Undulator Design Study and First Results}
\author{B.~Krasch$^1$, F.~Abusaif$^1$, T.~Arndt$^2$, N.~Glamann$^1$, A.~Grau$^1$, R.~Nast$^2$, D.~Saez de Jauregui$^1$\\
$^1$Institute for Beam Physics and Technology (IBPT), Karlsruhe Institute of Technology (KIT), Karlsruhe, Germany\\
$^2$Institute for Technical Physics (ITEP), Karlsruhe Institute of Technology (KIT), Karlsruhe, Germany
\thanks{bennet.krasch@kit.edu}}

\maketitle

\begin{abstract}
Undulators are X-ray sources that are widely utilised in advanced synchrotron radiation sources and free-electron laser facilities.
Due to sustainability and energy efficiency, the development focuses on small-scale, high-field, and especially compact undulators with short period lengths ($\leq$ 10\,mm) and narrow magnetic gaps ($\leq$ 4\,mm).
Therefore, high-temperature superconducting (HTS) tapes, which can provide a high critical current density and high critical magnetic fields at higher temperatures, are extensively used and investigated at the Karlsruhe Institute of Technology.
A new concept of superconducting undulators (SCUs) was introduced and further developed by laser-scribing a meander pattern into the superconducting layer to achieve a quasi-sinusoidal current path through the HTS tape.

Here we present the complete and final design of such a compact SCU, in particular our latest results regarding the design of the liner with its tapering in the warm region, as well as the cryogenic cooling concept.
\end{abstract}


\section{Introduction}
Insertion devices (IDs), such as undulators, are a crucial component for free-electron laser (FEL) facilities and advanced synchrotron radiation sources because they produce coherent photon beams with high brilliance and a broad spectral range.
In principle, an ID consists of an arrangement of alternating magnetic dipoles, guiding electrons on a sinusoidal path through the device.
The period length of the undulator is chosen in that way, that each time an electron emits synchrotron radiation due to deflection, and additionally  the emitted light interferes constructively with the radiation at all other deflection points along the undulator \cite{MOTZ_UNDULATOR}.
Up to now, most recent undulator technology is based on the permanent magnets, so called permanent magnet undulators (PMU).
By opening up the field towards superconducting materials by its commercial availability, superconducting undulators (SCUs) moved into the focus.
As the current standard, promising results were achieved by using NbTi.
However, Nb$_3$Sn is increasingly being used to generate higher magnetic fields.
With the latter, the handling and winding of the coils is more complex and more prone to faults, as special heat treatment is required for Nb$_3$Sn \cite{PRESTEMON_WIRES}.
In comparison to cryogenic PMUs, SCUs offer a higher magnetic peak field on symmetry axis for the same geometry (e.g. magnetic gap and period length) \cite{CASALBUONI_SCU20, CASALBUONI_SCUdevelopment, BAHRDT_VERGLEICH} and thus a higher tunability and brightness resulting in a wider range of applications at modern synchrotron light sources. 
This technology therefore holds great potential not only for current accelerators and beamlines, but also for future storage rings and linear accelerators such as FELs and laser plasma accelerators (LPAs).
As the demands for sustainability, energy efficiency and resource conservation lead to the realisation of more compact accelerator facilities, IDs must follow this trend.
But due to the natural extension of a NbTi wire cross-section, period lengths $\leq$ 8\,mm are not feasible.
That is where high-temperature superconductors (HTS) become a promising candidate. 
To meet these new requirements, new concepts have been developed to produce more compact undulators based on HTS \cite{KIM_HTS, CASALBUONI_HTS, CALVI_HTS, ZHANG_HTS}.
Current developments in SCU undulator technology are described here \cite{ZHANG_SCU, GRATTONI_SCU, CASALBUONI_SCU}.
To reach shorter period length, Prestemon et al. suggested to use non-insulated HTS tapes instead of wire and to structure the tape's surface with a meander pattern \cite{PRESTEMON_TAPE, PRESTEMON_TAPE2}.
Based on these considerations, T. Holubek \cite{HOLUBEK_TAPE} and A. Will \cite{WILL_TAPE} have developed different realisations and methods, both of which are further developed at KIT.
T. Holubek et al. designed a system using a 15\,m long structured HTS tape, wound in a manner that stacks 30 layers of these tapes to produce an alternating magnetic field. 
To conserve HTS material and reduce costs, A. Will et al. suggested soldering short individual structured tapes together in an accordion-like configuration.

In the following sections, the development of an HTS SCU based on structured tapes is presented.
We start with an overview of the manufacturing process of the structuring and the two stacking methods.
However, the focus is on the cooling concept, in particular on the integration of the magnetic coils in the cryostat and on the tapering for connection to a beam tube.

\section{Realisation of structured HTS tapes and stacks for a compact SCU}
\begin{figure}
    \centering
    \includegraphics[width=0.5\textwidth]{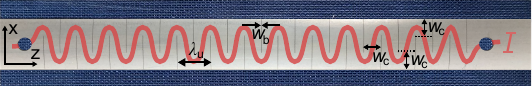}
    \caption{\textbf{Laser-structured section of a HTS tape.} Shown is the laser-structured meander pattern. The chosen parameters for the structuring section are: barrier width $w_\mathrm{b}=$ 25\, $\mu$m and channel width $w_\mathrm{c}=$ 4\,mm resulting in a period length $\lambda_\mathrm{u}=$ 8.05\,mm. There is a total of $N$= 12.5 periods. The holes on the left and right side are for the precise alignment of structured section to each other. The distance between the holes is 11\,cm. The assumed path of the electric current through the HTS tape is displayed in red. This structured pattern stays the same regardless of how the stacking process looks like.}
    \label{fig:single_tape}
\end{figure}

The HTS material used in the present work was purchased from the US company SuperPower Inc\footnote{The HTS wire specification can be found on the SuperPower Inc. website
at https://www.superpower-inc.com/specification.aspx.}.
The second generation (2G) HTS tape are in general 12\,mm wide and 50\,$\mu$m thick.
The 50\,$\mu$m thickness consists of a 2\,$\mu$m silver layer on the top, followed by a 1.6\,$\mu$m thick REBCO (rare earth barium copper oxide) HTS layer, a 0.2\,$\mu$m buffer stack, a 50\,$\mu$m substrate layer and another 1.8\,$\mu$m silver layer at the bottom.
The 20\,$\mu$m copper layer for stabilisation was omitted to keep the tape as thin as possible.
Instead of a 50\,$\mu$m substrate layer, also 30\,$\mu$m tape was used for a stack.
In the next step, the HTS tape is structured in a meander shape using high-energy laser pulses.
At the KIT Institute for Technical Physics (KIT-ITEP) laser-structuring is performed with an infrared Nd:YAG laser with a wavelength of 1030\,nm (model: TruMicro 5025, company: TRUMPF).
Figure \ref{fig:single_tape} shows such a structured area of an HTS tape.
The structured section is about 11\,cm long and has two holes to align it precisely with other sections.
In order to achieve the desired alternating arrangement of north and south poles, as typical for an undulator, two layers must be aligned in alternating current direction with a 180$^\circ$ phase shift between the layers. 
To achieve this type of stacking, two different approaches were tested and realised: a jointless and a soldered design.
Details regarding the laser unit, such as resolution and precision of the reel-to-reel system as well as the optical setup can be found in the publication by R. Nast et al. \cite{NAST_LASER}.
For more information on the parameters used to precisely cut at least the silver and HTS layer without bringing too much energy into the system to prevent it from overheating, see the publication by Krasch et al. \cite{KRASCH_MT}.

\subsection*{Jointless concept: single 15\,m long HTS tape}
T. Holubek et al. use a single 15\,m long HTS tape, which is structured in a meander shape according to the above process.
In the next step, the HTS tape is folded in the centre so that the structured surfaces face each other.
The HTS tape is then wound around a stainless steel yoke in such a way that the structured areas are precisely aligned.
The stainless steel yoke contains two pins for alignment that match the holes in the HTS tape.
In total, the 15\,m stacks 30 layers on top of each other.

\subsection*{Soldered concept: 30 single 25\,cm long HTS tapes}
\begin{figure}
    \centering
    \includegraphics[width=0.5\textwidth]{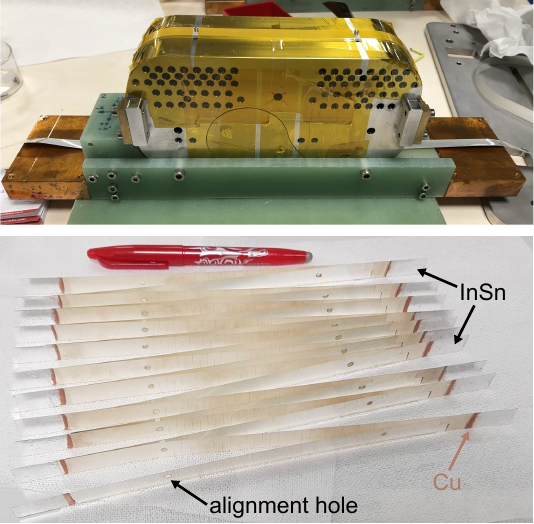}
    \caption{\textbf{Jointless and soldered concept.} The upper picture shows a stainless steel yoke around which a single 15\,m long structured HTS tape is wound. The alignment pins are later removed. The Kapton foil is used to fix the tapes to the yokes surface. On the left and right are copper plates displayed for the connection to the current leads. The lower picture shows 20 single structured HTS tapes which are soldered.}
    \label{fig:both_methods}
\end{figure}

The approach pursued by A. Will et al. focused on saving material and enabling the replacement of damaged tape.
The main difference to the approach of T. Holubek et al. is that instead of a single long structured HTS tape, 30 individual structured HTS tapes with a respective length of 25\,cm are connected to each other via a solder contact at the ends in the form of an accordion-like arrangement.
The special feature is that a cut-out is required at one end to enable the tape to be turned by 180° to allow the current to flow in the opposite direction.
For more information regarding the production process, especially the galvanisation process for the soldering contact, see \cite{KRASCH_MT}.
In contrast to the first method, the second method offers three advantages.
The first advantage is that there is a material saving of up to 75\%, as a large part of the 15\,m long HTS tape is only needed for the actual winding around the yoke and therefore remains unused for structuring.
The second and far greater advantage is that a "degraded" tape (due to handling or a manufacturing error) can generally be replaced.
With the first method, the 15\,m must always be disposed of as a whole.
It also saves costs.
The third advantage concerns handling.
50\,$\mu$m thick HTS tapes are easier to handle in short pieces and less likely to be broken than a 15\,m long HTS tape by undergoing the critical bending radius.
One disadvantage of the second method is the soldered contact between the individual single tapes, because it leads to an ohmic resistance.
To prepare the final stack, one starts to solder two tapes together, until 10 pairs are completed. After that two pairs are soldered together and so forth until 20 tapes are combined in a single stack.
The handling of 20 tapes in a stack is quite challenging.
Figure \ref{fig:both_methods} shows a comparison of the two methods for preparing.

\section{Magnetic Coil System}
\begin{figure}
    \centering
    \includegraphics[width=0.5\textwidth]{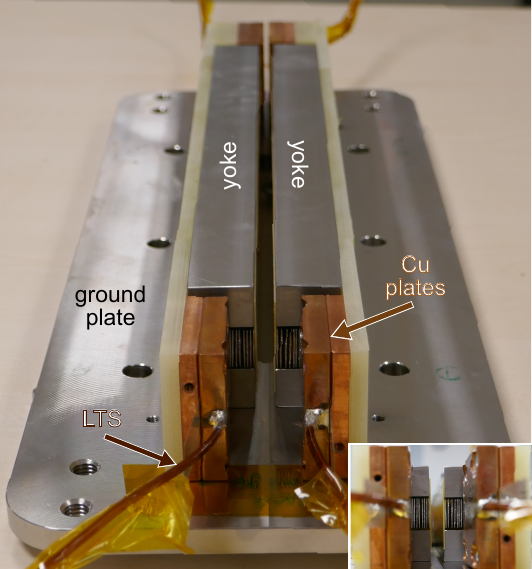}
    \caption{\textbf{Holding structure for the soldered stack.} Top view of the mounted yokes. Each yoke contains a stack of soldered HTS tapes and is mounted to the ground plate. The lowest HTS tape each is clamped in between two copper blocks to connect with the LTS current lead. On the opposite side the uppermost tapes are also clamped and the LTS wires are directly soldered together. This means there is only one circuit. The glass-reinforced plastic (GRP) sits on top of the stack and ensures that the top tape is flush with the surface of the yoke.}
    \label{fig:holding_structure_zigzag}
\end{figure}

In the jointless concept, each 15\,m long HTS tape is wound around its stainless steel yoke, see figure \ref{fig:both_methods} top picture.
In the soldered concept, the stainless steel yoke has a 12\,mm wide groove to host the stack and is flattened at the sides to provide space for the soldered contacts.
For both concepts, the two yokes are mounted on a holding structure (ground plate) with alignment holes to maintain the well-defined magnetic gap.
Figure \ref{fig:holding_structure_zigzag} shows as an example of the soldered concept.
The coils are then powered up to check that they can carry the current and remain superconducting.
The magnetic fields were measured at 4\,K and supplemented by simulations with the software OPERA (magneto-static TOSCA solver).
Detailed information can be found in the publications by D. Astapovych et al. \cite{ASTAPOVYCH_MESSUNGEN}and B. Krasch et al. \cite{KRASCH_MT}.

\section{Thermal Design}
\begin{figure}
    \centering
    \includegraphics{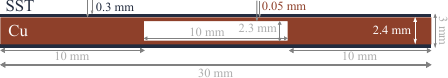}
    \caption{\textbf{Dimensions of the liner.} The liner is cuboid with a rectangular base area of 30\,mm wide and 3\,mm high and mainly made by copper (Cu) with a 0,3\,mm thin layer of stainless steel (SST) on top and bottom of the Cu block.}
    \label{fig:Liner}
\end{figure}

The thermal design for a compact HTS undulator is an important but challenging step towards its final realisation.
From the beginning of the design phase three boundary conditions were set:
\begin{itemize}
    \item First, the use of an electron beam chamber (liner) to absorb all beam heat loads to prevent heating the magnet coils.
    \item Second, the cooling of the undulator is refrigerant-free (e.g. no liquid helium or nitrogen) and therefore cryocoolers (CC) are implemented.
    \item Third, the tapering (transition from the geometry of the liner to the geometry of the accelerator beam pipe) takes place outside of the vacuum chamber at room temperature. This implies special thermal transition elements.
\end{itemize}
The cooling concept includes a \textbf{liner} for three main reasons: to absorb beam heat loads without heating the HTS tape, to reduce impedance effects on the electron beam, and to separate the different vacuum levels between the cryostat and beam tube.
However, it requires a larger magnetic gap, reducing the magnetic field, and adds cost.
The liner is cuboid in shape with a rectangular base area of 30\,mm wide and 3\,mm high, see Figure \ref{fig:Liner}.
It is mainly made of copper with a residual-resistivity ratio (RRR) $\geqslant$ 100.\footnote{The size of the RRR is a measure of the purity of a material. The higher the ratio, the fewer impurities the material has.}
A 0.3\,mm stainless steel layer for stabilisation and connection to the vacuum chamber is added at the top and bottom.

\begin{figure}
    \centering
    \includegraphics{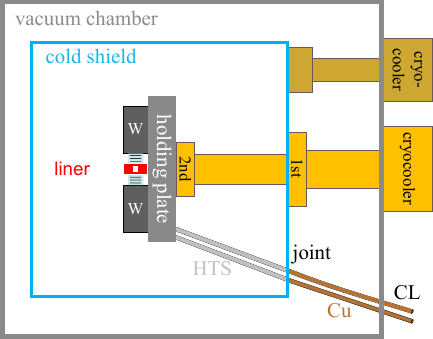}
    \caption{\textbf{Thermal design of the vacuum chamber in beam direction.} In the centre, the rectangular liner is shown in red. The 30 layer stacks of HTS tapes are above and below it and are fixed to the yokes (W). The yokes are mounted to the holding plate which is mounted directly on the 2\textsuperscript{nd} cold stage of the two-stage CC. This arrangement is located inside the cold shield, which is connected to the 1\textsuperscript{st} stage of the two-stage CC and to the single-stage CC. The copper current leads (CL) are shown here in simplified form without feed-through connectors. }
    \label{fig:Vakuumkammer}
\end{figure}

According to the above specifications, Figure \ref{fig:Vakuumkammer} shows the cooling concept of the \textbf{cryostat} of the compact HTS undulator in simplified form.
The design uses a two-stage Gifford-McMahon CC providing 1.5\,W at 4\,K.
The coils are mounted on the 2\textsuperscript{nd} stage to guarantee 4\,K for maximum current through the tapes.
The liner connects to the cold shield via its front and rear side.
The cold shield is connected to both CCs.
The reason for the second CC (single-stage RDK-400B with 50\,W at 30\,K) is that the provided cooling capacity only by the Gifford-McMahon would lead to a temperature of 70\,K of the liner according to simulations.
Table \ref{tab:heat_loads}
\begin{table}
        \centering
           \caption{Heat loads to the cold stages of the Gofford-McMahon CC.}
        \begin{tabular}{lcc}
         \toprule
             Source of heat load & 1\textsuperscript{st} stage (W) & 2\textsuperscript{nd} stage (W)\\
         \midrule
             Current leads (CL) & 34 & 0.15 \\
             Instrumentation CL & 0.05 & - \\
             Beam heat load & 0.1 & - \\
             Thermal radiation & 10 & 0.01 \\
             \textbf{in total} & \textbf{44.15} & \textbf{0.16} \\
         \bottomrule
        \end{tabular}
        \label{tab:heat_loads}
\end{table}
gives an overview about the different types of heat loads and their contribution.
The publication by B. Krasch et al. \cite{KRASCH_MT} provides more information on thermal design, in particular the (beam) heat loads and their contribution, and thermal simulations of the of the liner and the coils temperature distributions using the COMSOL Multiphysics software.

\begin{figure}[h!]
    \centering
    \includegraphics[width=0.4\textwidth]{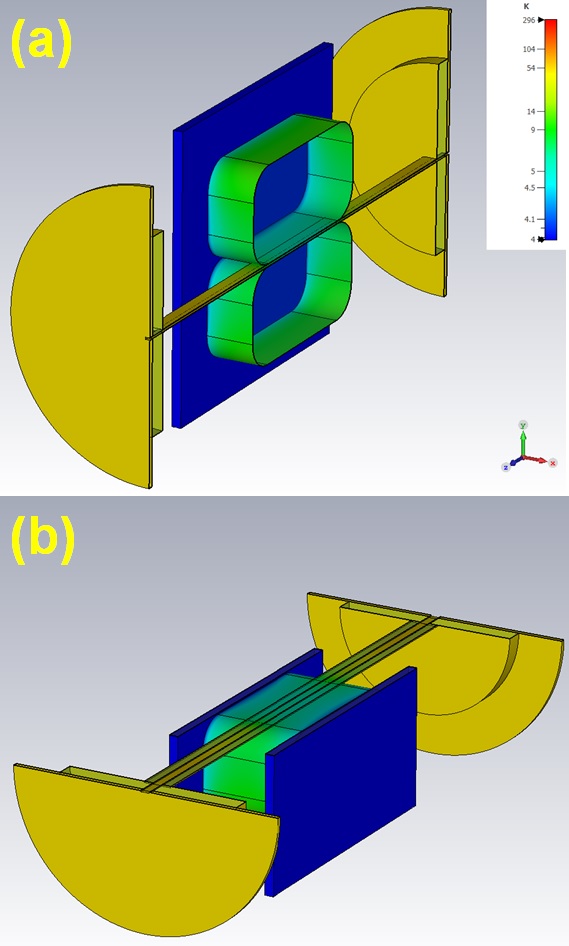}
    \caption{\textbf{Temperature distributions in the liner and its peripheral components shown with cutting planes of (a) y-z
    and (b) z-x, respectively.} The yokes are shown in green, with the liner between them.
    The holding structure is shown in blue and the thermal copper transitions at the ends of the liner are shown in orange. }
    \label{fig:Cha_IPAC23_fig3}
\end{figure}

\begin{figure}[h!]
    \centering
    \includegraphics[width=0.4\textwidth]{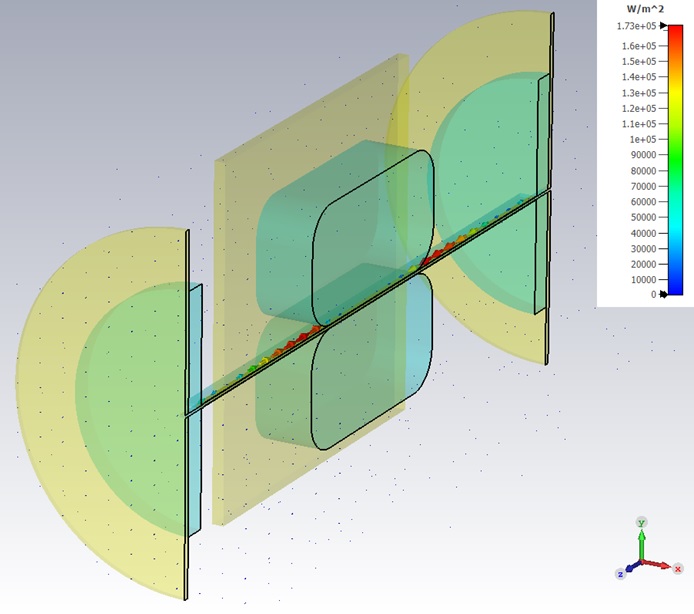}
    \caption{\textbf{Distribution of heat flow density at the liner.} }
    \label{fig:Cha_IPAC23_fig4}
\end{figure}

In contrast to B. Krasch et al. who thermally characterised the isolated cryostat with liner and the magnets, the following simulations show the temperature distribution for the \textbf{thermal transition modules}.
Information regarding the design of the thermal transition module are part the next section.
The thermal transition elements are between the liner and the cold shield.
The CST Studio Suite® was used to simulate the thermal distribution of the liner at cryogenic temperatures and including the thermal transitions to check whether they are suitable at all.
Figure \ref{fig:Cha_IPAC23_fig3} shows the temperature distribution in the liner and its peripheral components when 4\,K and 50\,K were applied to the Cu plates and the thermal shields, respectively.
Steady-state thermal simulations showed a temperature range of 5.5\,K to 7.5\,K around the beam axis, which is sufficient for HTS tapes.
The total heat intake by conduction in the liner was calculated to be approximately 0.2\,W.
Heat flow density distribution is shown in Figure \ref{fig:Cha_IPAC23_fig4}.
The publication by Cha et al. \cite{CHA_THERMISCH} contains further information on the thermal simulation and the mechanical analyses.
\begin{figure*}[h!]
    \centering
    \includegraphics[width=\textwidth]{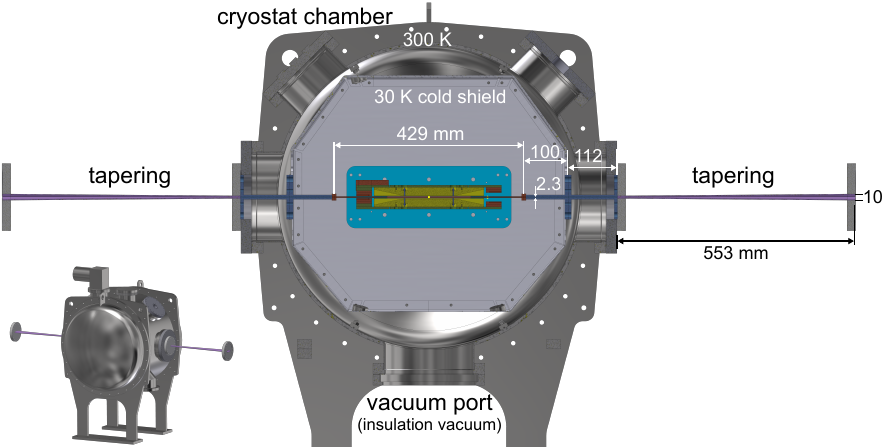}
    \caption{\textbf{Final vacuum design of the compact HTS undulator.} Starting from the outside, the cryostat chamber, which is at room temperature (300\,K), is located in the centre of the figure.
    Further inside is the cold shield, which is cooled to 30\,K by two CCs, see Figure \ref{fig:Vakuumkammer}.
    The holding structure for the magnetic coils (HTS stacks and yokes) is located in the centre of the cryostat in turquoise.
    The liner is located between the two yokes and has a total length of 429 \,mm.
    The liner is connected to the cold shield via the first connection  (100\,mm) and to the flange on the outside of the cryostat via a second connection (112\,mm).
    The tapering starts outside the cryostat and ends after 553\,mm at the flange of the accelerator beam pipe.
    This gives a length of 979.5\,mm from the centre of the cryostat to the beam pipe flange, or 1959\,mm from beam pipe flange to beam pipe flange. Flanges for feedthroughs (measurement instrumentation) are available in the upper area of the cryostat. A turbomolecular pumping station generating a vacuum of at least $10^{-6}$\,mbar is mounted to the flange at the bottom. The complete, closed cryostat is shown in the lower left corner. }
    \label{fig:cryostat}
\end{figure*}   
In summary, it can be said that the thermal simulations and calculations show that the designed cooling concept fulfils the requirements mentioned at the beginning.
In the next section, the final vacuum design is presented with regard to tapering and the thermal transition modules.

\section{Vacuum Design: Transition Modules and Tapering}
\begin{figure}[!]
    \centering
    \includegraphics{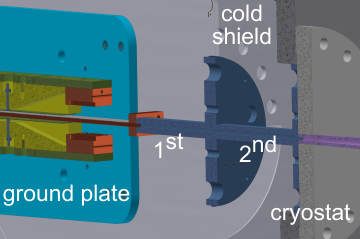}
    \caption{\textbf{Transitions from the liner to the cryostat.} An enlarged section of the inside of the cryostat is shown. The 1\textsuperscript{st} transition (100\, mm long) starts with a copper block at the end of the liner and ends at the thermal shield. The 2\textsuperscript{nd} transition (112\,mm long) connects the cold shield to the outer wall of the cryostat.}
    \label{fig:transition}
\end{figure}

After explaining the fabrication of the HTS stacks and the thermal concept, this section describes the entire vacuum system, focusing on the thermal transition modules and the connection to the accelerator beam tube.
Figure \ref{fig:cryostat} shows the cross section along the beam axis of the cryostat with all the components involved.
A major challenge was the design of the appropriate interfaces and transitions for the liner (design of the liner, see Figure \ref{fig:Liner}).
According to Figure \ref{fig:transition},
the liner ends with a copper block.
Starting with another copper block, the first transition element begins (labelled 1\textsuperscript{st} in Figure \ref{fig:transition}) and connects the end of the liner to the cold shield.
\textbf{hier}
This transition corresponds to the transition module simulated above.
This chamber has the same internal height of 2.3\,mm as the liner, so there is no step that could create an impedance on the electron beam.	
The second transition element (labelled 2\textsuperscript{nd} in the figure \ref{fig:transition}) connects the 30\,K cold shield to the flange on the outside of the cryostat and has a length of 112\,mm and an internal height of 2.3\,mm. 
This results in a total length of 212\,mm of transition.

\begin{figure}[!]
    \centering
    \includegraphics{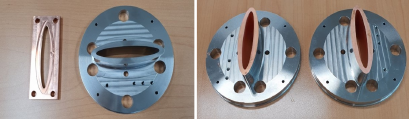}
    \caption{\textbf{Thermal transition modules.}}
    \label{fig:thermalmodule}
\end{figure}

Another challenge lies in the design and fabrication of appropriate thermal transition modules.
On one hand, the goal is to make these modules as thin as possible to minimise heat input; on the other hand, experience has shown that thicknesses of just a few tens of micrometers are insufficient.
Due to thermal expansion, such thin modules tend to deform and break.
Based on these experiences, a complete new thermal module was not only designed but also fabricated, installed, tested, and simulated for a previous experiment where the beam heat load to a cold bore was measured, see \cite{CHA_TRANSITION}.
These new thermal modules can be seen in Figure \ref{fig:thermalmodule}.
In the first step, an oxygen-free high conductivity (OFHC) copper block and a stainless steel component are manufactured using the wire cutting method.
The stainless steel part features a 5\,mm thick transition and a mounting flange.
The inner surfaces of this module are smoothly polished with diamond powder.
In the next step, a thin copper layer (approximately 10\,$\mu$m) is applied to the inner surfaces of the stainless steel component to significantly reduce the influence of mirror charges caused by resistive wall heating, as shown in the lower right picture of Figure \ref{fig:thermalmodule}.
In the third and final step, the copper block is firmly joined with the stainless steel component.
This ensures that the mechanical stress caused by thermal contraction does not lead to leakage or breakage under vacuum conditions.
High mechanical accuracy of the transition modules is essential for the successful integration with the existing components in the cryostat.
The installation and subsequent successful tests have shown that the fabricated modules function and perform as intended.
For more details on the fabrication process of the modules, installation considerations, and commissioning results, refer to the publication by Cha et al. \cite{CHA_TRANSITION}.
It is important to say that these thermal transition modules were designed for an another experiment, but due to the successful realisation (fabrication and testing in an experiment) transition modules for the compact HTS undulator  will be fabricated based on these modules by Cha et al. \cite{CHA_TRANSITION} while closely following the design presented above.

\begin{figure}
    \centering
    \includegraphics{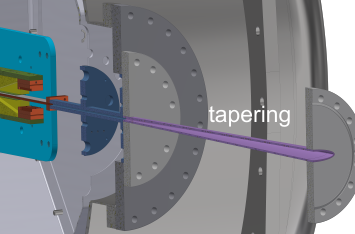}
    \caption{\textbf{Tapering from cryostat to accelerator beam pipe.} The picture shows the previous components with the outer wall of the cryostat. The tapering between the cryostat flange and the accelerator beam pipe flange is shown in purple.}
    \label{fig:tapering}
\end{figure}

\begin{figure}
    \centering
    \includegraphics[width=0.5\linewidth]{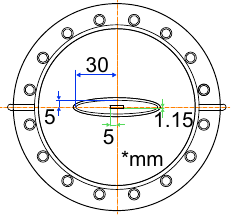}
    \caption{\textbf{Dimensions of the elliptical beam pipe.}}
    \label{fig:beam pipe}
\end{figure}

In the final step, the transition to the accelerator's beam pipe is considered.
It has a length of 553\,mm.
Together with the length of the transition modules, 765\,mm is the distance between the end of the liner and the beginning of the beam pipe.
Since the geometry of the accelerator beam pipe (elliptical) is different to the geometry of the liner (rectangular), a corresponding tapering is required, see Figure \ref{fig:tapering}.
The elliptical shape, see Figure \ref{fig:beam pipe},
was chosen as a typical accelerator beam pipe. 
Nevertheless, the tapering can be easily adjusted to any beam pipe geometry.

The final vacuum design for a compact SCU based on laser-structured HTS tapes has been successfully completed.
Based on previous experiments, the main design parameters for the thermal transition modules are known.

\section{Summary and Outlook}
The field of high-temperature superconductors (HTS) for undulators is still relatively young compared to NbTi wound undulators.
However, HTS offers advantages in terms of energy efficiency and durability, and are of great importance for the construction of more compact accelerators.
We presented two methods for manufacturing HTS undulators with a meander-shaped structure.
We described and discussed the corresponding cooling concept for such compact SCUs, the final integration into a cryostat, and the connection (tapering) to the accelerator's beam pipe.
Special transition modules are needed to lead the electron beam chamber from the inside of the cryostat to the outside, leading to the final vacuum and cryogenic design shown.

The next steps include the manufacturing of the thermal transition modules and a final assembly.
This represents an important step towards the realisation of compact undulators based on laser-structured HTS tapes.

\section{Acknowledgements}
This work was supported by the German Federal Ministry of Education and Research (BMBF) under the grant no. 05K19VK1.

\end{document}